\documentclass{article}
\usepackage{spconf,amsmath,graphicx,multirow,cite,url,booktabs}
\usepackage{comment}
\usepackage{subfigure}
\usepackage[para]{threeparttable}
\usepackage[noend]{algpseudocode}
\title{Detection and evaluation of human and machine generated speech in spoofing attacks on automatic speaker verification systems}
\name{Yang Gao$^*$, Jiachen Lian$^*$, Bhiksha Raj$^+$, Rita Singh$^+$}

\address{
  $^*$Electrical and Computer Engineering, $^+$Language Technologies Institute\\
  Carnegie Mellon University\\
  Pittsburgh, PA, USA - 15213}

\begin{document}
%\ninept
%
\maketitle
\begin{abstract}
Automatic speaker verification (ASV) systems utilize the biometric
information in human speech to verify the speaker's identity. The
techniques used for performing speaker verification are often
vulnerable to malicious attacks that attempt to induce the ASV system
to return wrong results, allowing an impostor to bypass the system and
gain access. Attackers use a multitude of spoofing techniques for this,
such as voice conversion, audio replay, speech synthesis, etc. In
recent years, easily available tools to generate deepfaked audio have
increased the potential threat to ASV systems. In this paper, we
compare the potential of human impersonation (voice disguise) based
attacks with attacks based on machine-generated speech, on black-box and white-box ASV systems. We also study countermeasures by using features that capture the unique
aspects of human speech production, under the hypothesis that machines
cannot emulate many of the fine-level intricacies of the human speech
production mechanism. We show that fundamental frequency
sequence-related entropy, spectral envelope, and aperiodic parameters
are promising candidates for robust detection of deepfaked speech
generated by unknown methods.
\end{abstract}
\section{Introduction}
\label{sec:intro}

% ASV 系统会受到攻击；
Automatic speaker verification (ASV) systems utilize the biometric information in human speech to verify the identity of a speaker by matching it with the information present in a database (which is also derived from speech samples). Such systems are also vulnerable to malicious attacks where the attacker tries to provide fake biometric information to fool the ASV systems. 
% 攻击的种类有很多，比如
There are many spoofing methods in use by attackers nowadays, including direct human impersonation of the target, machine assisted-speech generation such as voice conversion (VC), customized and manipulated text-to-speech synthesis (TTS) system outputs, etc. 
% 除此之外，因为深度学习技术的发展，生成的攻击也变得越来越强。
With the advancement of deep learning techniques, especially with advancements in generative models such as generative adversarial networks \cite{gao2018voice} and wavenet models \cite{vanwavenet, 2018tacotron2}, the quality of synthetic speech is getting much closer to natural speech \cite{2018tacotron2}. Attacks carried out using synthetic speech generated by these methods pose serious threats to ASV systems. As a first step, in this paper, we compare the threats from impersonation attacks with the synthetic speech attacks and establish that deep synthesized fakes are in fact the most dangerous attacks for ASV systems. 

\begin{figure}[t]
 \centering
 \includegraphics[width=\linewidth]{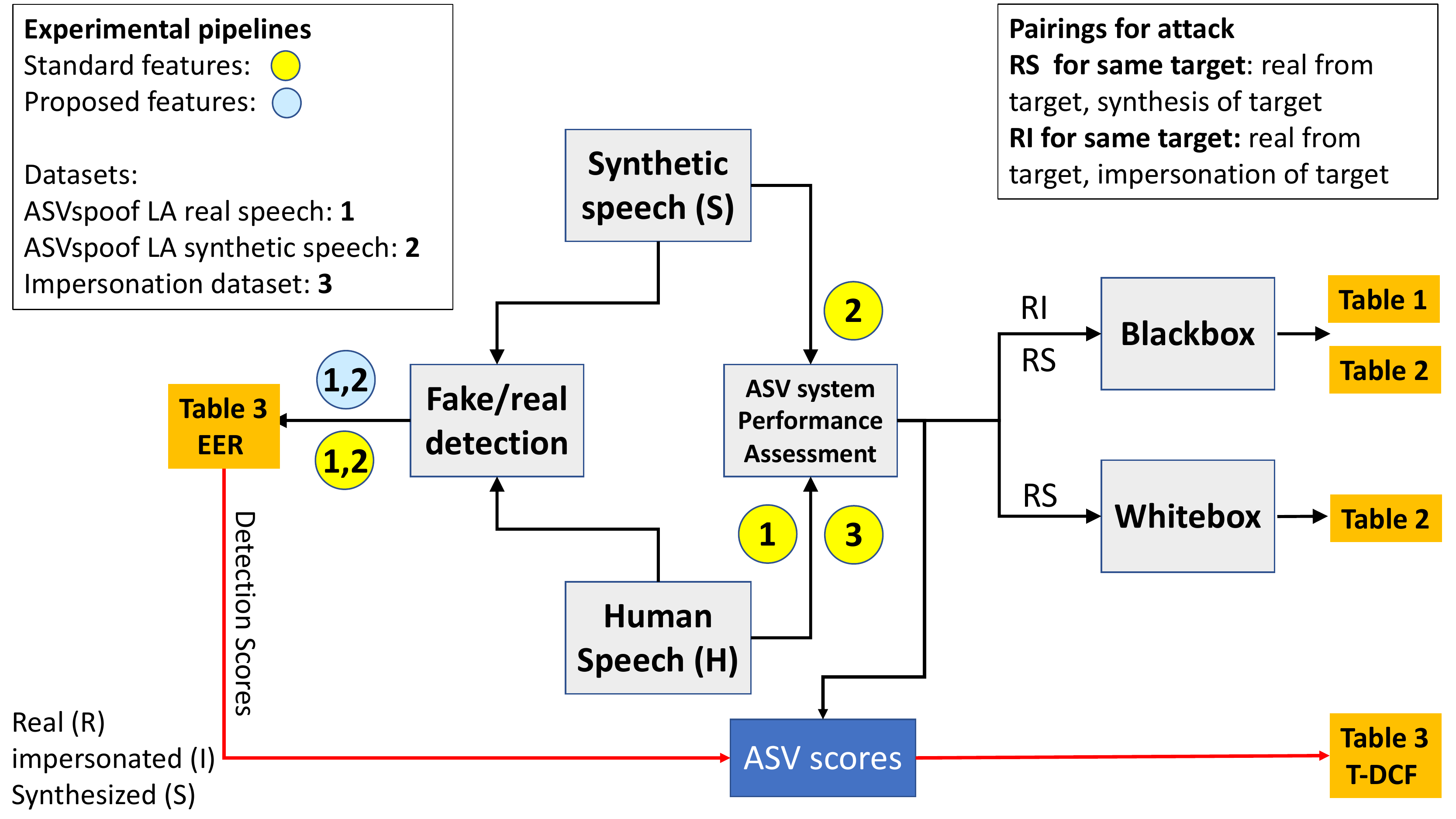} 
 \label{fig:diagram}%
 \vspace{-0.5cm}
 \caption{Diagram for the experimental setups}
 \vspace{-0.2cm}
\end{figure}

It is therefore important to be able to distinguish between fake/synthesized and human-generated speech. This is the broader goal of our paper. However, we do not focus on mere feature selections as \cite{balamurali2019toward, kamble2020advances, yu2017dnn}, which would be influenced by the dataset choices, models and training procedures. Instead, we start with the hypothesis that machine-generated speech is too consistent in many respects, and machines are unable to emulate the finer level variations found in naturally produced speech signals. In other words, because of the complexity of the human speech production mechanism, human speech has a greater degree of inconsistency than machine-generated speech. We devise experiments to investigate a select set of features that we believe capture some intricacies of human-generated speech in a manner that machines cannot.
To verify the correctness of our hypothesis, we must not only evaluate the features directly on detecting fake and real speech signals but also establish this through applications such as ASV systems, that must then, by the use of these features in the countermeasures, become more impervious to malicious attacks. We propose several speech-generation-related features and verify them to be effective in improving the overall performance of the detection model and ASV systems. The experiments diagram of this paper is shown as in Fig. \ref{fig:diagram}.

\subsection{Prior work}

% 基于此，有一些很好的数据集和challenge已经有所研究，比如ASV challenge系列和FOR数据集。
A number of approaches of varying success have been proposed in the literature to detect fake speech to increase the security of ASV systems against spoofing attacks. For example, the long-running ASVspoof challenge \cite{wu2015asvspoof, kinnunen2017asvspoof, todisco2019asvspoof} has raised wide efforts in fake speech spoofing attack countermeasures on ASV systems. 
The main focus of the challenge, however, has been to rank spoof detection countermeasures, and not to carry out an in-depth evaluation of the ASV systems' performance under attacks. 
Another significant problem with the spoof detection study is that, with the rapid evolution of deepfake generation methodologies, the sheer variety of attacks that an AVS system may be subject to is also rapidly increasing. Detection models trained on a specific provided dataset synthesized using a limited set of methods are not likely to generalize well to newer types of fake/generated audio \cite{chen2020generalization}.  
In ASVspoof2019, for example, detection algorithms \cite{ASVres2019deep, das2019asvlong, yang2019significanceasv, zeinali2019detectingASV} that work very well on training datasets are often found to perform much worse on evaluation sets that have been produced using attack techniques not present in the training data. The detection performance on the evaluation sets could be an indicator of the generalization capacity of those proposed algorithms as one purpose of the challenge. 

To include the latest deep-learning speech synthesizers, \cite{FoR} provides a synthetic speech dataset called Fake or Real (FoR), improving variety of the deepfake speech data for this purpose. We have, in fact, also used it effectively in the context of this work. 

% Data from synthetic speech detection challenges can still be useful for the evaluation of ASV attack countermeasures if used judiciously. For example, \cite{FoR} provides a synthetic speech dataset called Fake or Real (FoR), improving variety of the data for this purpose. We have, in fact, used it effectively in the context of this work. 

\section{Comparing Human and Machine generated speech in Spoofing attacks}

As a first step, in this paper, we compare the threats from impersonation attacks with the synthetic speech attacks and establish that deep synthesized fakes are in fact the most dangerous attacks for ASV systems. 

\subsection{Datasets} \label{Datasets}

% ! write dataset seperately in details. Especially about the CID dataset/ 

For our experiments we use four datasets: the logical access (LA) of ASVspoof 2019 dataset \cite{todisco2019asvspoof}, the VoxCeleb dataset \cite{NAGRANI2020101027}, the FoR dataset \cite{FoR} and our own collected impersonation dataset (CID). 
% a short description for each of the above four datasets

The CID dataset is collected from the performances of expert impersonators on YouTube, and segmented carefully to only keep the speech segments corresponding to target speakers. All impersonators in the CID dataset are professionals mimicking political figures for amusement, collecting from TV shows and talk shows on YouTube. The dataset comprises 1091 utterances of genuine speech from both the impersonators and the political figures and 981 utterances of impersonated speech. The data are further segregated into paired sets, each pair containing an impersonator's real speech and target/mimicked speech, or the target's real speech and the speech produced for the same target by an impersonator. These are indicated in Table \ref{tab:impersonation-asv-eer}, in which pairs that originate from the same speaker are called positive pairs, and pairs from different speakers are called negative pairs. 

As shown in Table \ref{tab:impersonation-asv-eer}, we have two sets of positive pairs and four sets of negative pairs. They are 19086 positive pairs of same target speaker's real utterances (R), 14844 negative pairs of different target speaker's real utterances (RI), 3382 positive pairs of same impersonator's impersonations for different target (IAB), 37554 negative pairs of target and impersonation pair (TI), 1988 negative pairs of different impersonator's real utterances (IRAB) and 28080 negative pairs of target/impersonator's real utterance pair (IRT). 

For synthetic data, ASVspoof2019 dataset contains logical access data and physical access data. In this study, we only use the logical access data which contains machine-generated speech using multiple text-to-speech synthesis and voice conversion methods. The logical data has 2580 bonafide utterances and 22800 synthetic utterances from 20 speakers in the training set; 2548 bonafide utterances from 20 speakers and 22296 spoof utterances from 10 speakers in the development set \cite{todisco2019asvspoof}. The evaluation set contains 7355 bonafide utterances from 67 speakers and 63882 spoof utterances from 48 speakers. The spoof audios are generated using unseen spoofing algorithms intentionally, aiming to give insights of the generalization performance of the proposed countermeasure models. In order for a general ASV system to evaluate this dataset, we generate 4914 bonafide positive pairs and 4914 negative pairs for each attack (A07-A19) from original evaluation set. We also generate 15970 positive pairs and 15970 negative pairs to evaluate the overall attacking ability over all attacks.

To best evaluate the threats of different attacks, we train ASV systems under unconstrained recording and speaking conditions (essentially data-in-the-wild). For this, we use the VoxCeleb dataset, which is a large scale publicly dataset containing millions of utterances collected from unconstrained speech samples \cite{NAGRANI2020101027}. It has many speakers and millions of utterances under different recording conditions. This can be effectively used to evaluate the potential of any given ASV methodology to generalize to unseen speakers and unconstrained conditions \cite{NAGRANI2020101027, chung2020in}. 

\subsection{Analyzing performance under attacks on black-box and white-box ASV systems} \label{Implementation details}

The ASV model we use is proposed by Joon Son Chung, et al. \cite{chung2020in}, which applies the Thin ResNet-34\cite{NAGRANI2020101027} as backbone, and Self-attentive Pooling(SAP)\cite{cai2018exploring} as aggregation strategy. This model, when trained with short-time Fourier transform (STFT) spectrogram of Voxceleb, generalizes extremely well to unconstrained conditions as shown by the low EER of real utterance pairs, mentioned earlier in this section. 
% We call this our \textbf{baseline ASV model}. To evaluate the (different) features in our work, a minor modification of the baseline model is made by substituting the adaptive pooling layer for average pooling before SAP,  so that the model can incorporate features of different sizes. We call this latter model the \textbf{modified ASV model}.

The \textbf{black-box} ASV system is pretrained with the VoxCeleb dataset.
STFT, mel-frequency cepstral coefficients (MFCCs), aperiodic parameters(AP) and spectral envelope(SP) are used as input features to this model. The original input audios comprise segments of 2sec duration. We use the same STFT feature as in \cite{chung2020in}\cite{NAGRANI2020101027}. MFCC feature is computed from 16kHz sampled signals: which comprise 13 cepstral coefficients, to which first and second-order derivatives respectively are concatenated, making the feature dimensionality 39. (AP and SP are not the focused of this section and will be further discussed in Section.\ref{sec:features} and Section.\ref{sec:Discussion})

The \textbf{white-box} model is trained with the ASVspoof 2019 data, as a multi-class classifier for speaker identification. We make small modification on the initial ASVspoof2019 training set by assigning each spoofed utterance an identity which uniquely incorporates both speaker and attack. There are 20 speakers and 6 types of attack in the ASVspoof2019 LA training set, meaning that there are 120 "spoofed identities". Thus our modified training set contains 140 identities. We call these \textit{ASVspoof training identities} (ASVTIs). 

% In the second stage, we use the four black-boxes as pre-trained models to train the ASVTIs using Angular Prototypical loss function\cite{chung2020in}, where M is 2. For comparison purposes, in an additional stage, we apply the modified model to train ASVTIs with Spec/MFCCs1/MFCCs2 while not using any pre-trained models. 
% During the evaluation process, all 4 blackbox models and 7 whitebox models are tested on ASVspoof evaluation pairs (as described in \ref{Datasets}) to compute ASV EER. We evaluate the attack ability of different generation approaches using our ASV system,  and find the most threatening generation methods. 

\subsubsection{Impersonation attacks}

\begin{table*}[t]
\vskip 0.5 \baselineskip
\caption{EERs of impersonation attacks to the ASV under black-box scenario}
% \vspace{-0.7cm}
\label{tab:impersonation-asv-eer}
\begin{center}
\scalebox{0.95}
{
\begin{threeparttable}[t]
\begin{tabular}{lc | cccccccc}
\toprule
\multicolumn{2}{c}{} & \multicolumn{8}{c}{ASV EER\%}\\
\hline
\multicolumn{2}{c}{Impersonation Data} & \multicolumn{1}{c}{R\tnote{1}+RI\tnote{2}} & \multicolumn{1}{c}{IAB\tnote{3}+RI} & \multicolumn{1}{c}{R+TI\tnote{4}} & 
\multicolumn{1}{c}{IAB+TI}  & 
\multicolumn{1}{c}{R+IRAB\tnote{5}} & \multicolumn{1}{c}{IAB+IRAB} & 
\multicolumn{1}{c}{R+IRT\tnote{6}} & 
\multicolumn{1}{c}{IAB+IRT} \\
\hline
\midrule
blackbox & \begin{tabular}{@{}c@{}@{}c@{}c@{}} VoxCeleb2(STFT) \\ VoxCeleb1(STFT) \\ VoxCeleb1(MFCC) \\VoxCeleb1(AP)\\VoxCeleb1(SP) \end{tabular} 
& \begin{tabular}{@{}c@{}} 1.71\\4.16\\17.21\\39.75\\53.58 \end{tabular} 
& \begin{tabular}{@{}c@{}} 13.30\\16.41\\22.09\\41.27\\49.42\end{tabular} 
& \begin{tabular}{@{}c@{}} 11.42\\14.95\\22.01\\44.89\\54.36\end{tabular}
& \begin{tabular}{@{}c@{}} 43.52\\42.02\\48.77\\45.46\\50.36\end{tabular} 
& \begin{tabular}{@{}c@{}} 4.86\\4.74\\9.22\\45.58\\55.04\end{tabular} 
& \begin{tabular}{@{}c@{}} 17.76\\15.90\\26.80\\46.06\\51.11\end{tabular} 
& \begin{tabular}{@{}c@{}} 5.21\\5.06\\8.86\\41.65\\53.76\end{tabular}
& \begin{tabular}{@{}c@{}} 19.45\\15.64\\20.11\\42.94\\49.43\end{tabular}  \\
\bottomrule 
\hline
\end{tabular}

\begin{tablenotes}
\item[1] {\footnotesize R: Same target speaker's real utterance pair ($+$, \#19086)}
 \item[2] {\footnotesize RI: Different target speaker's real utterance pair ($-$, \#14844)}
 \item[3] {\footnotesize IAB: Same impersonator, impersonations for different target pair ($+$, \#3382)}
 \item[4] {\footnotesize TI: Target and impersonation pair ($-$, \#37554)}
 \item[5] {\footnotesize IRAB: Different impersonator's real utterance pair ($-$, \#1988)}
 \item[6] {\footnotesize IRT: Target and impersonator's real utterance pair ($-$, \#28080)}
 \item[7] {\footnotesize The model pre-trained with VoxCeleb2 dev set using Spectrogram feature}
 \hrule
\end{tablenotes}
\end{threeparttable}
}

\end{center}
% \vspace{-0.4cm}
\end{table*}

% ! Make here more clear, solve the reviewer point

Our black-box evaluations on impersonation attacks use the CID dataset. We run several experiments to evaluate the dataset's attacking potential. The results are shown in Table \ref{tab:impersonation-asv-eer}. The model that is pretrained on VoxCeleb2 is able to verify open-set speakers best and gives $1.71\%$ EER for target speakers' real utterances (positive and negative pairs R + RI); This pretrained model can be seen as a black box ASV under open-set evaluation. 

From our tests, we observe that combining the impersonation/target pairs (TI) with the positive pairs from real speakers (R) improves the speaker verification EER to $11.42\%$, which indicates that professional impersonation can fool the ASV system to a certain extent, although it is still ineffective in most attacks. 
The group with real speaker positive pairs (R) and negative pairs (IRAB) built from the real voices of different impersonators has a low EER of $4.86\%$, showing that the pre-trained ASV system is indeed generalized to verify unseen target speakers and cross impersonators pairs. 

The IAB is the same impersonator mimicking different targets. IAB + RI has an EER of $13.3\%$, showing that even if the same speaker tries to impersonate different targets, their utterances are mostly considered as the same speaker, although still having some capacity to fool the ASV system. Note that this EER value gets significantly larger to $22.09\%$ with less generalized models, such as the VoxCeleb1 pre-trained model. This indicates that impersonation from professional impersonators is still threatening to some ASV systems. The IAB + IRAB set has comparable EERs as the IAB + RI, showing that the IRAB pairs are valid, also indicating the true differences between the impersonator's real voices.  And the IAB + TI gives an EER of $43.52\%$. This high EER 
comes from the formation of this evaluation set. Different from other sets, both the negative pairs and positive pairs can be seen as the spoofing attacks because the professional impersonator could impersonate different target to a certain extent, which makes the 'positive' pairs negative in nature. Therefore, both the positive pairs of IAB and the negative pairs of TI are hardest cases, also showing by the EERs of their combination with the R and RI. 

The R + IRT corresponds to positive pairs for the real voice utterances of the same targets and negative pairs of impersonator's real voice with the targets' real voice. The $5.21\%$ EER shows that the impersonator's real voices are indeed not similar to the targets' voices.   

The overall results show that while mimicry from amateur impersonators is reported to not succeed in fooling ASV systems in previous research \cite{kinnunen2019mimicry, vestman2020voiceMimicryJ}, mimicries rendered by professional impersonators still poses threats to a certain extent. 

% Traditionally, it is interesting to note that human impersonation is considered to pose relatively lower levels of threat to ASV systems. It is harder for a human to emulate low-level intricacies of someone else's speech. For impersonation, fortunately for mimics, simply emulating the coarser perception-related features of speech is enough to give the impression of the target speaker. Some prior studies, e.g., \cite{kinnunen2019mimicry, vestman2020voiceMimicryJ} have found that mimicry from amateur impersonators does not succeed in spoofing the tested ASV system. However, in these studies, the mimicries rendered were from untrained impersonators. This is likely to be the case with most malicious attackers, who may be able to mimic only very few of their target speakers well enough to carry out large-scale voice spoofing attacks on ASV systems.  
% 我们想知道，如果是专业的模仿者，会不会对ASV造成困扰
% However, there is still the possibility that more dangerous attackers may use professional impersonations to break into ASV systems. It is, therefore important to extend our investigations to human impersonation attacks.

\begin{figure}%
% \vspace{-0.5 cm}
\centering
\subfigure[][]{%
\label{fig:ex3-a}%
% \vspace{-0.8 cm}
\includegraphics[height=1in]{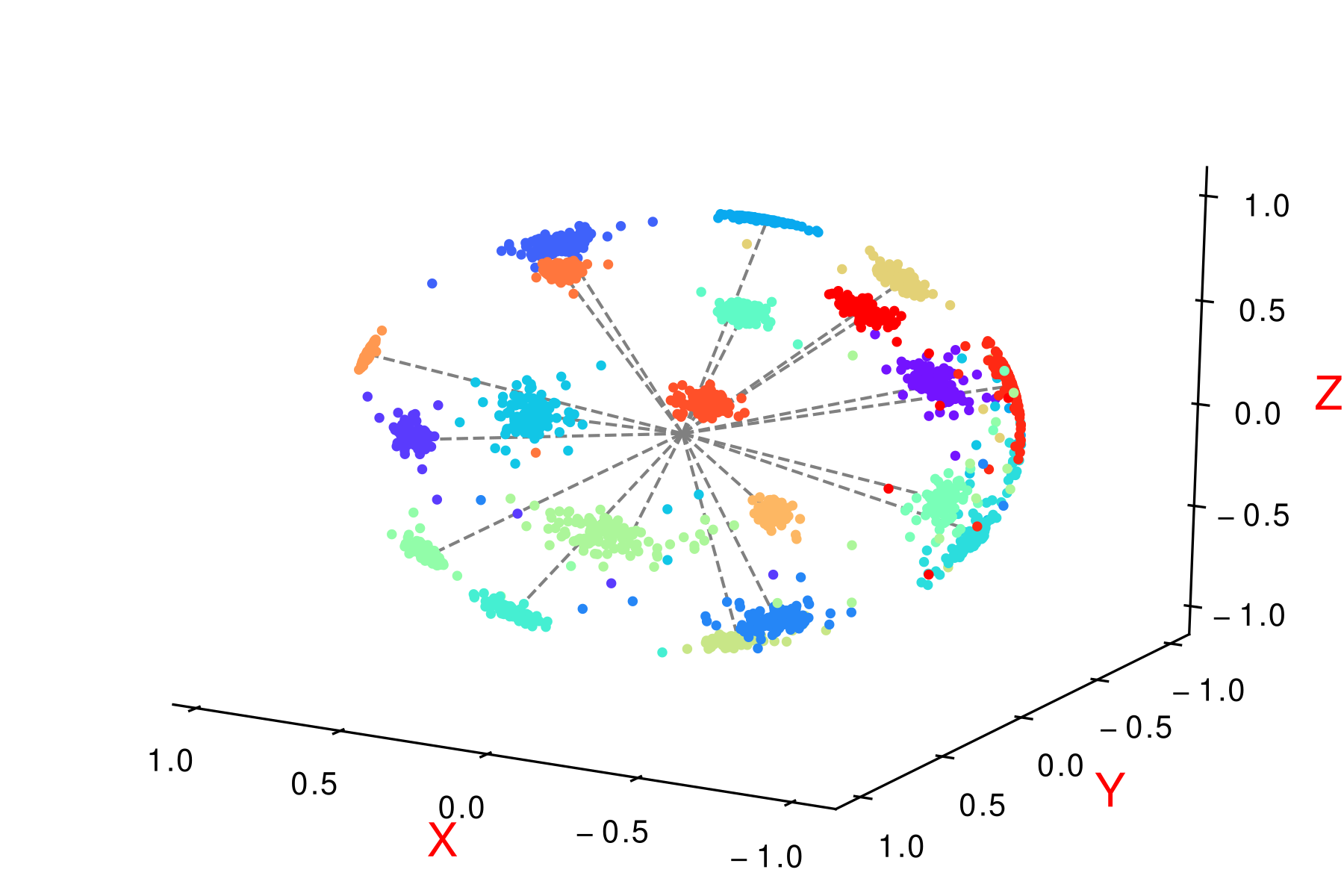}}%
% \vspace{-0.5 cm}
\hspace{8pt}%
\subfigure[][]{%
\label{fig:ex3-b}%
\includegraphics[height=1in]{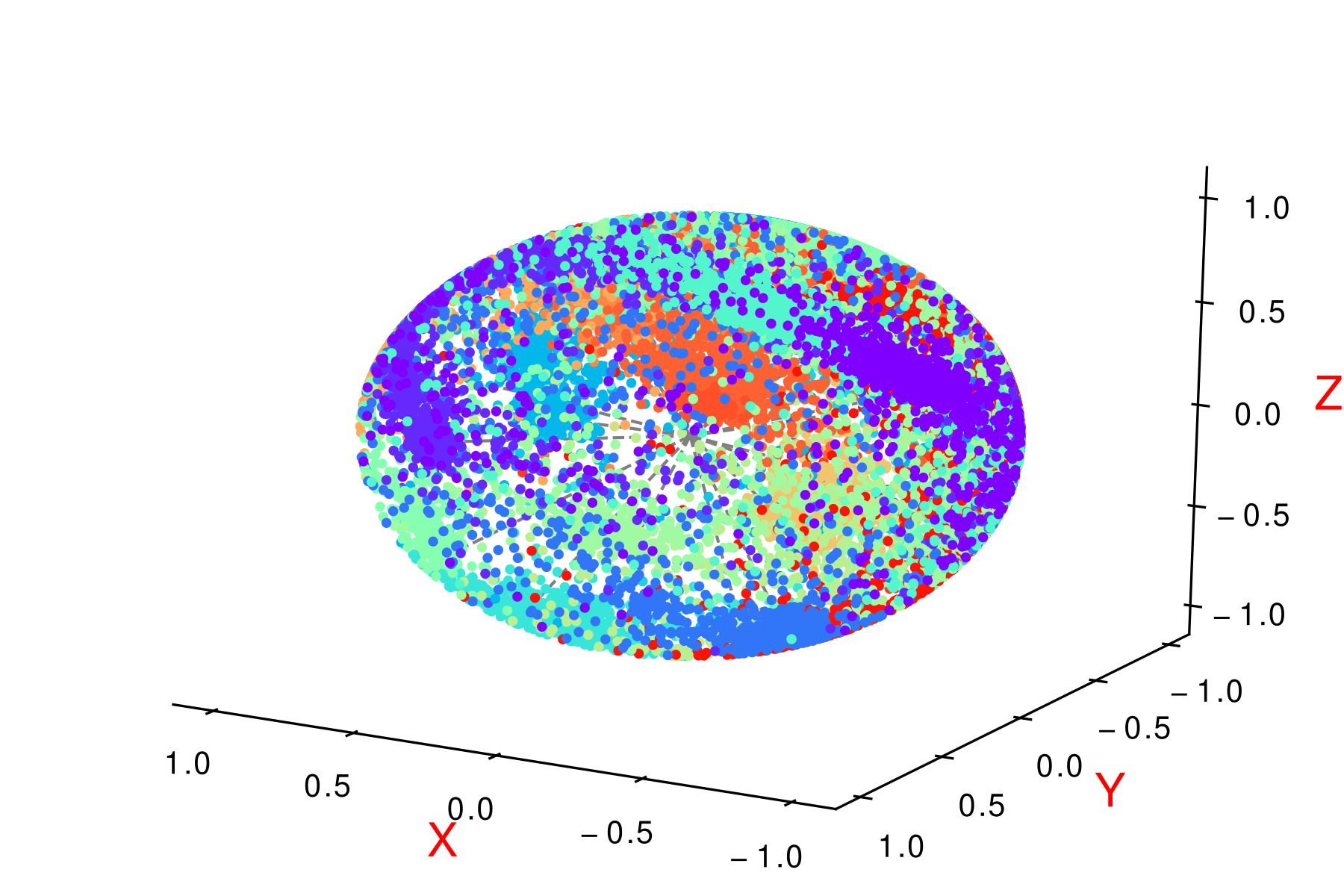}} \\
\vspace{-0.4 cm}
\subfigure[][]{%
\label{fig:ex3-c}%
\includegraphics[height=1in]{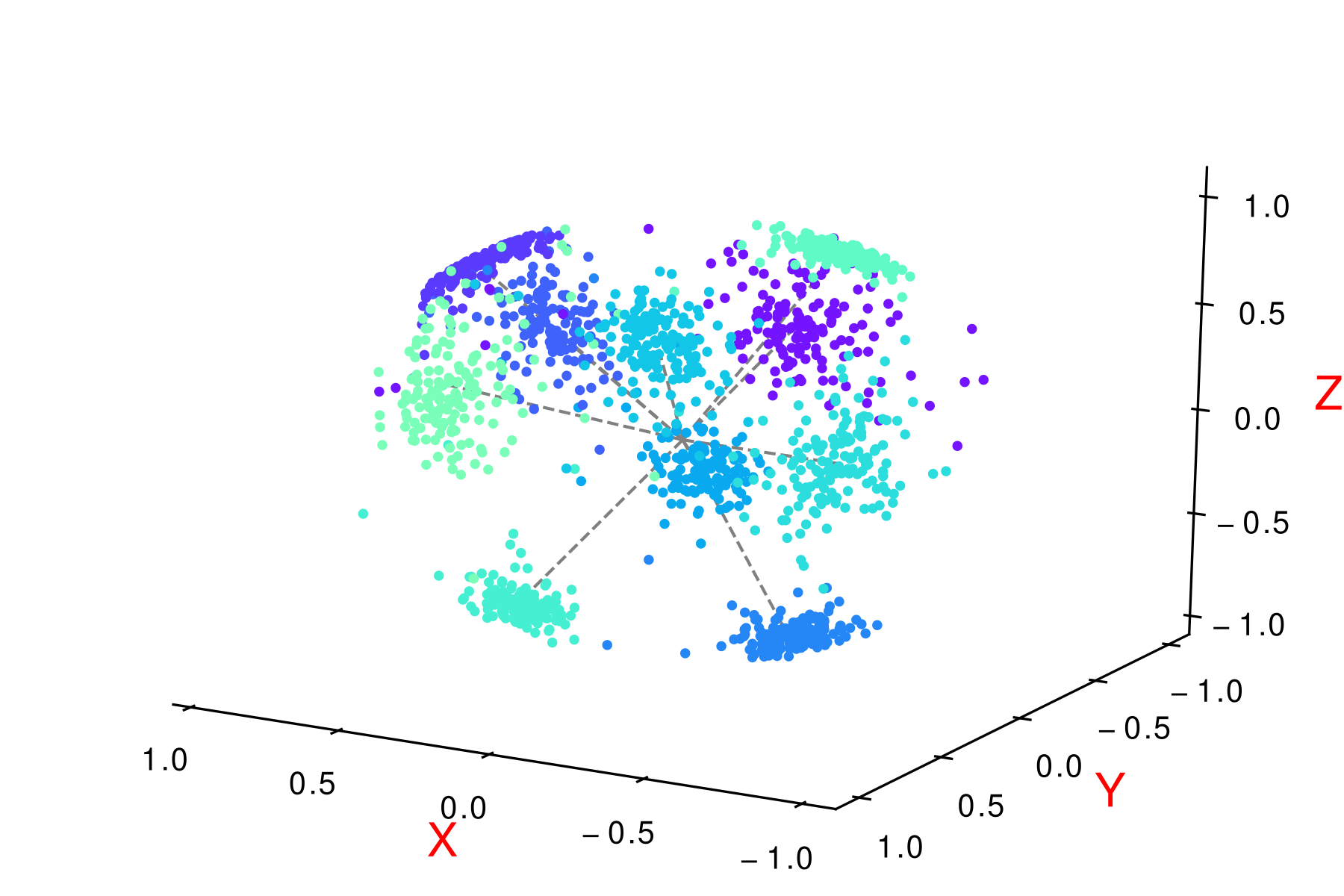}}%
\hspace{8pt}%
\subfigure[][]{%
\label{fig:ex3-d}%
\includegraphics[height=1in]{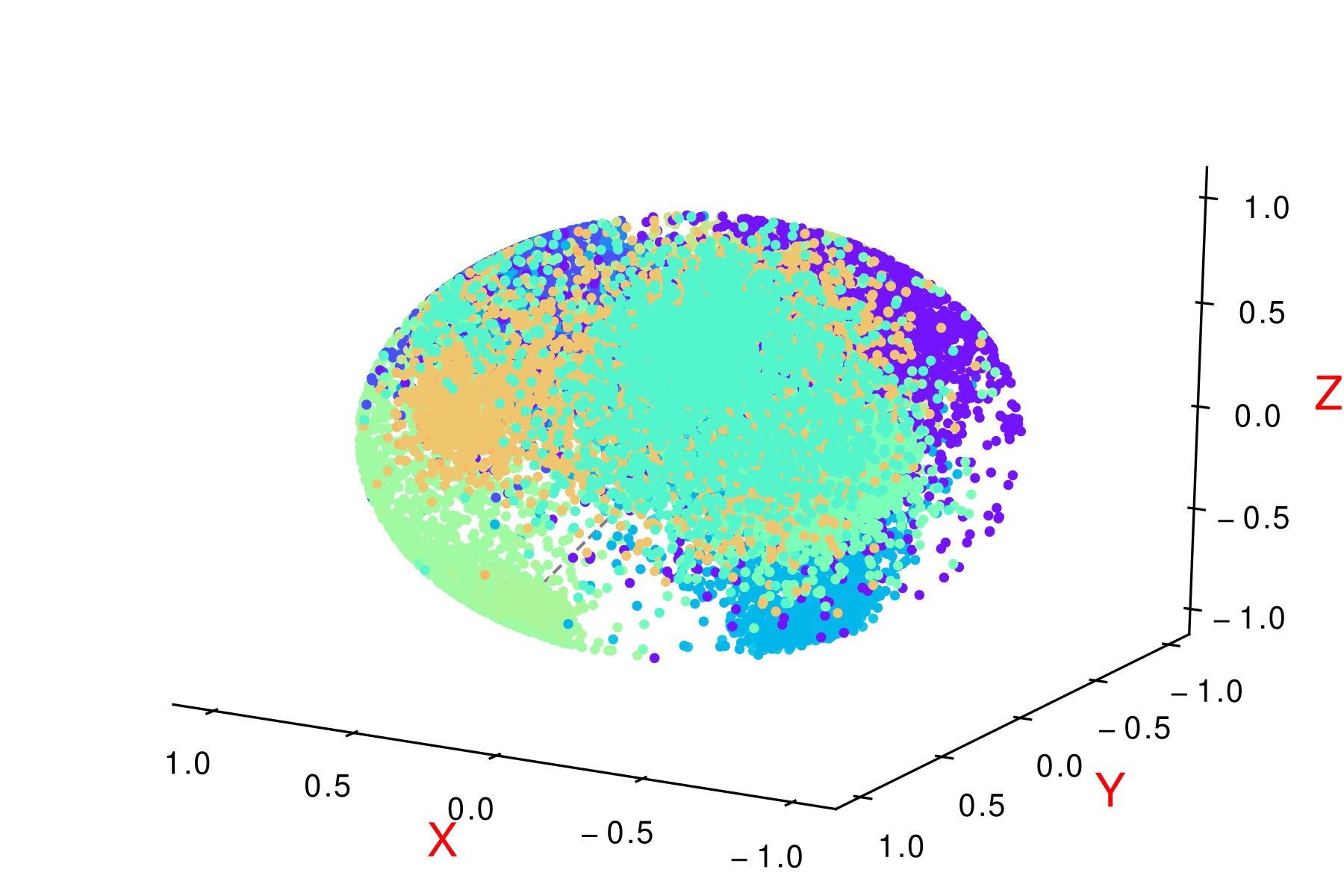}}%
% \vspace{-0.2 cm}
\caption[Feature Space]{
\subref{fig:ex3-a} 20 bonafide speakers in training set;
\subref{fig:ex3-b} 20 bonafide and spoof speakers in training set;
\subref{fig:ex3-c} 10 bonafide speakers in development set;
\subref{fig:ex3-d} 10 bonafide and spoof speakers in development set.}%
% \vspace{-0.6 cm}
%\captionsetup{belowskip=0pt}
\label{fig:ex3}%
% \vspace{-0.8 cm}
\end{figure}

\subsubsection{Synthetic speech attacks}
% \vspace{-0.2 cm}

\begin{table*}[!t]
\caption{EERs of evaluation set for ASVspoof2019 LA under black-box and white-box scenarios}
% \vspace{-0.7cm}
\label{tab:asv-eer}
\vskip 0.5 \baselineskip
\begin{center}
% \begin{small}
% \resizebox{\columnwidth}{!}
\scalebox{0.74}
{
\begin{threeparttable}[t]
\begin{tabular}{lc | cccccccccccccccc}
\toprule
\multicolumn{2}{c}{} & \multicolumn{13}{c}{ASV EER\%}\\
\hline
\multicolumn{2}{c}{Attack} & \multicolumn{1}{c}{A07} & \multicolumn{1}{c}{A08} & \multicolumn{1}{c}{A09} & \multicolumn{1}{c}{A10}  & \multicolumn{1}{c}{A11} & \multicolumn{1}{c}{A12} & \multicolumn{1}{c}{A13} & \multicolumn{1}{c}{A14} & \multicolumn{1}{c}{A15} & \multicolumn{1}{c}{A16} & \multicolumn{1}{c}{A17} & \multicolumn{1}{c}{A18} &
\multicolumn{1}{c}{A19} &
\multicolumn{1}{c}{ALL\tnote{3}}\\
\hline
\midrule
blackbox & \begin{tabular}{@{}c@{}@{}c@{}c@{}} VoxCeleb2(STFT)\\VoxCeleb1(STFT) \\ VoxCeleb1(MFCC)\\VoxCeleb1(AP)\\VoxCeleb1(SP) \\Todisco et al., \cite{todisco2019asvspoof}\end{tabular} 
& \begin{tabular}{@{}c@{}} 34.03\\27.93\\45.12\\39.89\\51.47\\59.68 \end{tabular} 
& \begin{tabular}{@{}c@{}} 23.20\\25.30\\28.89\\24.92\\50.73\\40.39\end{tabular} 
& \begin{tabular}{@{}c@{}} 5.70\\11.01\\16.02\\31.63\\53.51\\8.38\end{tabular}
& \begin{tabular}{@{}c@{}} 48.51\\47.77\\45.01\\38.66\\51.46\\57.73\end{tabular} 
& \begin{tabular}{@{}c@{}} 37.37\\37.36\\48.88\\21.93\\53.97\\59.64\end{tabular} 
& \begin{tabular}{@{}c@{}} 43.42\\44.77\\45.09\\43.05\\49.52\\46.18\end{tabular} 
& \begin{tabular}{@{}c@{}} 23.67\\30.93\\38.13\\30.99\\54.51\\46.78\end{tabular}
& \begin{tabular}{@{}c@{}} 40.45\\43.33\\35.06\\28.25\\49.92\\64.01\end{tabular} 
& \begin{tabular}{@{}c@{}} 43.14\\40.91\\43.01\\42.21\\51.39\\58.85 \end{tabular} 
& \begin{tabular}{@{}c@{}} 50.51\\43.36\\46.04\\45.55\\50.66\\64.52\end{tabular} 
& \begin{tabular}{@{}c@{}} 4.99\\7.65\\11.07\\31.56\\52.71\\3.92\end{tabular}
& \begin{tabular}{@{}c@{}} 7.10\\10.83\\25.64\\31.92\\49.18\\7.35\end{tabular} 
& \begin{tabular}{@{}c@{}} 11.26\\13.97\\25.02\\39.27\\54.97\\14.58\end{tabular} 
& \begin{tabular}{@{}c@{}} 21.42\\22.03\\25.66\\35.55\\50.60\\-\end{tabular} 
\\\hline
whitebox & \begin{tabular}{@{}c@{}@{}c@{}c@{}} ASVSpoof(STFT)\tnote{1}\\ASVSpoof(MFCC) \\ASVSpoof(AP)\\ASVSpoof(SP)\\ \textbf{VoxCeleb2+ASVSpoof(STFT)} \\VoxCeleb1+ASVSpoof(STFT) \\ VoxCeleb1+ASVSpoof(MFCC)\\VoxCeleb1+ASVSpoof(AP) \\ VoxCeleb1+ASVSpoof(SP) \end{tabular} 
& \begin{tabular}{@{}c@{}} 2.33\\7.12\\38.93\\50.97\\\textbf{1.16}\\1.21\\4.99\\22.04\\50.24 \end{tabular} 
& \begin{tabular}{@{}c@{}} 2.65\\5.08\\32.46\\49.94\\\textbf{2.31}\\2.63\\4.51\\17.77\\44.30 \end{tabular} 
& \begin{tabular}{@{}c@{}} 3.75\\8.12\\32.59\\40.07\\\textbf{0.77}\\1.75\\1.99\\28.13\\40.51 \end{tabular} 
& \begin{tabular}{@{}c@{}} 47.56\\39.76\\42.37\\49.75\\\textbf{43.42}\\45.85\\37.28\\33.68\\49.21 \end{tabular} 
& \begin{tabular}{@{}c@{}} 40.89\\28.99\\38.29\\49.25\\\textbf{27.62}\\17.40\\19.02\\37.78\\48.82 \end{tabular} 
& \begin{tabular}{@{}c@{}} 47.59\\49.01\\43.28\\52.04\\\textbf{41.23}\\45.69\\45.08\\37.20\\50.45 \end{tabular} 
& \begin{tabular}{@{}c@{}} 37.01\\33.81\\37.02\\52.30\\\textbf{15.46}\\20.84\\33.18\\19.72\\48.74 \end{tabular} 
& \begin{tabular}{@{}c@{}} 29.09\\19.04\\33.96\\51.03\\\textbf{34.48}\\25.85\\15.92\\\textbf{6.50}\\48.62 \end{tabular} 
& \begin{tabular}{@{}c@{}} 35.48\\41.39\\41.12\\51.74\\\textbf{36.26}\\25.41\\33.65\\33.16\\49.43 \end{tabular} 
& \begin{tabular}{@{}c@{}} 4.09\\9.08\\49.06\\51.99\\\textbf{6.63}\\4.66\\6.01\\44.43\\50.86 \end{tabular} 
& \begin{tabular}{@{}c@{}} 12.07\\18.00\\40.05\\41.49\\\textbf{1.26}\\2.24\\11.30\\33.25\\33.63 \end{tabular} 
& \begin{tabular}{@{}c@{}} 28.61\\16.47\\34.57\\46.16\\\textbf{5.75}\\8.24\\11.44\\32.33\\42.10 \end{tabular} 
& \begin{tabular}{@{}c@{}} 1.88\\2.09\\44.53\\45.78\\\textbf{0.68}\\0.73\\2.98\\41.01\\38.06 \end{tabular}  
& \begin{tabular}{@{}c@{}} 22.24\\15.99\\39.25\\42.08\\\textbf{11.99}\\13.35\\14.94\\32.45\\36.79 \end{tabular} \\
\bottomrule 
\hline
\end{tabular}
 \begin{tablenotes}

 \item[1] The model trained directly with ASVTIs\tnote{4} using Spectrogram feature
 \item[2] The model pre-trained with VoxCeleb1 dev set using Spectrogram(blackbox) and subsequently trained with ASVTIs
 \item[3] Evaluation on general pairs as described in \ref{Datasets}, indicating overall EER
 \item[4] As defined in \ref{Implementation details}
\end{tablenotes}
 
\hrule
\end{threeparttable}
}
\end{center}
% \vspace{-0.7cm}
\end{table*}

To further understand the attacks of synthetic speech generated from different methods, we perform extensive ASV evaluations on the ASVspoof evaluation set under the black-box and white-box conditions. The evaluation set contains attack methods from A07 to A19 which are different voice conversion or speech synthesis techniques \cite{wang2019asvspoof}.

Comparing human-generated attacks and machine generated attacks, as in the black-box scenarios for both cases, we found that the general attack ability of the machine generated speech are stronger than the human-generated attacks, as shown in Table.\ref{tab:impersonation-asv-eer} and Table.\ref{tab:asv-eer}. For the blackbox of VoxCeleb2 trained using STFT, most of the synthetic attacks have a EER of over $20\%$, much higher than the EER of impersonation attacks (IAB). 

As is shown in Table \ref{tab:asv-eer}, A09/A17/A18/A19 are relatively weaker attacks showing lower ASV EER$\%$ in STFT/MFCC-based black-boxes, which is consistent with the results given by \cite{todisco2019asvspoof}. These attacks are generated through waveform generators such as waveform filtering and spectral filtering, which may be simpler methods compared to hard cases using neural vocoders. Most of attacks tend to be more dangerous for MFCC-based black-box than STFT-based black-boxes.   
% White-boxes result in improvement of overall EER and over most EERs of single attacks, compared to their corresponding black-boxes. 
Also for STFT and MFCC, EERs for most attacks are lower under the white-box scenario, compared to attacks on black-boxes with the same dataset and feature settings. However, it does not result in much improvement of EER for A10, A12, and A15, which are generated by \textbf{neural waveform models}, indicating increased threat from deepfakes.

%A14

In conclusion, the feature robustness ranks as 'STFT $>$ MFCC' with finetuning under white-box scenario. This conclusion is consistent with our hypothesis that features that capture information about prosodic nuances are more robust under attacks for ASV systems. 

In Figure. \ref{fig:ex3}, we draw the embedding features from the bottleneck layer of the white-box ASV model on a sphere. When spoofed utterances are introduced, it is not easy to discriminate the embeddings anymore, which indicates their threats to the ASV system.

\section{Strategy for establishing the goodness of features for fake speech detection}
\label{sec:features}

The threatens shown above from the deepfakes indicates the urgent needs for fake speech detection to assist the anti-spoofing capacity of ASV systems. 
To do so, our final goal is to find robust features for detecting fake speech, especially deepfakes. To re-iterate, our hypothesis is that features that capture the fine-level nuances of human speech from a speech-production perspective are likely to be able to effectively help distinguish between real and fake speech. In addition, they are also likely to improve the performance of countermeasures that are used for thwarting ASV spoofing attacks carried out through synthetic speech.

% However, the robustness of features in this context is not easy to establish. We must show through simple experiments that they can indeed distinguish between real and fake speech with high accuracy, and at the same time, they \textit{increase the robustness of systems that verify speech}, i.e., make them more resilient to spoofing attacks.
% However, we must \textit{also} show that at the same time, they \textit{increase the robustness of systems that verify speech}, i.e., make them more resilient to spoofing attacks. 
% The latter must essentially follow from the former. In this paper, our goal is to show that it does, and based on the fulfillment of these two criteria, to show that the features we select are indeed good for fake speech detection. 
% To re-iterate, 
% Our hypothesis is that features that capture the fine-level nuances of human speech from a speech-production perspective are likely to be able to effectively help distinguish between real and fake speech. In addition, they are also likely to improve the performance of countermeasures that are used for thwarting ASV spoofing attacks carried out through synthetic speech. 

\subsection{Human voice-production based features} \label{3.1}

% must rephrase 
In the production of speech, there are several sources that are either aperiodic or periodic that generate acoustic energy in the vocal tract. The aperiodic sources are aspiration generated at the glottis, friction generated in the vocal tract, and transient bursts from the rapid release of complete constrictions. The periodic source is the vibration of the vocal folds that creates periodic energy at the glottis. Identifying and quantifying these various sources has several applications in speech coding, speech recognition, and speaker recognition \cite{deshmukh2005use}.
% Aperiodic sources include aspiration, generated at the glottis; friction, generated further forward in the vocal tract; and transient bursts produced by the rapid release of complete constrictions. The periodic source in speech is created by the vibration of the vocal folds creating periodic energy at the glottis. 

Synthetic utterances generated by deep generative systems lack specific aspects of naturalness. One notable example is that of prosody. While we do have high quality and plain prosody TTS datasets, these are far from perfect. This is likely to make prosody a promising candidate for our work. Prosody is partially represented through variations in the fundamental frequency (F0) of the speech signal. In addition, \textbf{features that capture prosody variations} are the F0 sequence, spectral envelope, and spectral aperiodicity. We evaluate all of these in our work. Our hypothesis is that features that capture the fine-level nuances of human speech from a speech-production perspective are likely to be able to help distinguish between real and fake speech effectively. Besides, they are also likely to improve the performance of countermeasures that are used for thwarting ASV spoofing attacks carried out through synthetic speech. For example, as shown in Fig.3, the spectral envelope information of fake speech lacks natural transition and nuances, consistent with our hypothesis that the synthetic utterances may lack some aspects of naturalness.

In the vocal production process of a human, the fundamental frequency we refer to is the natural frequency of the vibration of the vocal cords. A specific nuance we can leverage is (known from prior literature) that the larynx can be approximated a nonlinear dynamic system, and the vocal folds can be approximated to coupled oscillators that are theoretically capable of an infinite number of different vibration patterns. However, these are persistently in a perturbed state. In vocal acoustics, perturbation typically refers to a deviation from an expected regularity in vocal-fold vibration. No biological system can produce truly periodic oscillations, and some instantaneous fluctuation can always be expected \cite{coan2007handbook}. \textbf{Features that capture such instant-to-instant perturbations} are the well-studied 
\textit{jitter} and \textit{shimmer} measurements, which gauge the cycle-to-cycle variations in frequency and amplitude of the speech signal, respectively. It is expected that the information captured by jitter and shimmer may be differently ``enacted'' in machine-synthesized speech (if at all). We thus choose also to evaluate these features in our work.

\begin{figure}[t]
\centering
\includegraphics[width=\linewidth]{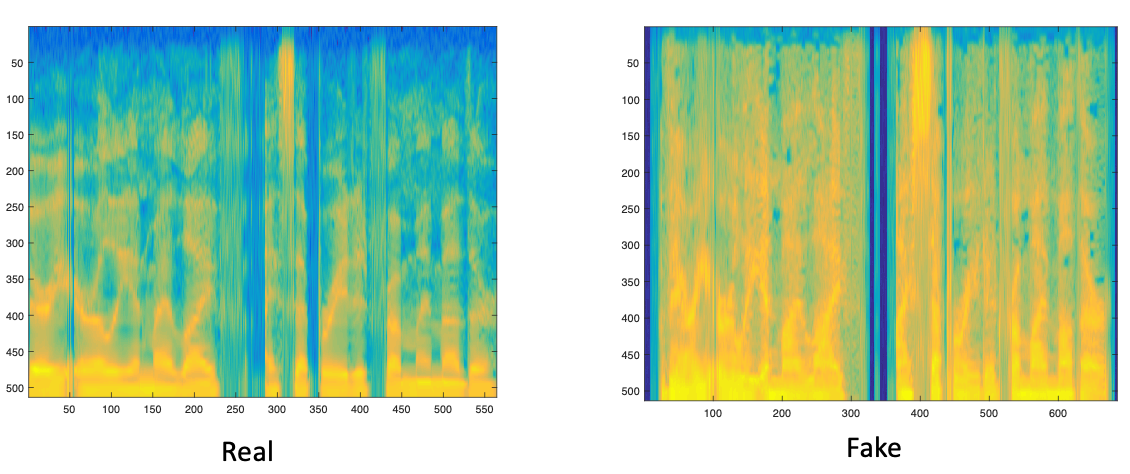} 
\label{fig:sp-vis}
\vspace{-0.5cm}
\caption{The same text utterance ("Very early in my life, I separated from my mother." )'s spectral envelope (log scale) of real speech contains natural transition and nuances while the fake speech does not. And the short pause is unnaturally sharp in fake speech.}
 % \vspace{-0.6cm}
\end{figure}

% This is indicated in \cite{NAGRANI2020101027}, wherein VGGVox, a VGG based model is proposed which gives an equal error rate (EER) of 2.87\% on Voxceleb1 dataset. In \cite{chung2020in},  a prototypical network together with angular prototypical loss is proposed for similar purposes, which gives an EER of 2.21\% on VoxCeleb1. Both VGGVox and the prototypical model of \cite{chung2020in} generalize well, and have the potential to be used to extract meaningful features ASV under unconstrained conditions.

\subsection{Analyzing robustness of speech-production motivated features} 

We are now in a position to analyze the robustness of speech production motivated features for detecting fake speech and improving the robustness of ASV systems against synthetic speech-based attacks. In fact an ASV countermeasure model that evaluates verification performance (based on t-DCF \cite{kinnunen2018tDCF} and EER measurement) automatically consolidates and verifies both goals. For reasons explained earlier, we choose to use jitter, shimmer, and other features that capture F0 variations. 

For experiments with jitter and shimmer, we only use the utterance-wise average jitter and shimmer values (extracted using Praat \cite{praatShimmer}), which may not be the best way to use such transient information from speech signals. Nevertheless, we build a three-layer MLP as a countermeasure model that uses these features. In our implementation, we set the F0 range to be within 75-500 Hz. The results show a 31\% EER on the development set, showing that even simple aggregates of these features (the average across an utterance in this case) already make a positive difference to performance.

For experiments with aperiodic and spectral envelope signal features, we verify the spoofing countermeasures in performance improvements. We use the \textbf{detection model} that is modified from the residual net architectures proposed in \cite{ASVres2019deep}. To evaluate the proposed features, we do not focus on fine-tuning parameters and use five residual blocks compared to the 9-11 blocks in \cite{ASVres2019deep} for all input features. We set the kernel to be of different sizes to accommodate the dimensionality requirements of the spectral envelope and aperiodic information extracted using WORLD \cite{morise2016world}. In our evaluation of these features, from Table \ref{tab: ap-sp}, the EERs are similar for the dev and eval set using these features alone. We can also see that the fusion of aperiodic information and spectral envelope with MFCC or spectrogram or Constant Q cepstral coefficients (CQCC) \cite{todisco2017cqcc} features can improve the detection performance as evaluated by EER and the joint performance with ASV evaluated by the t-DCF \cite{kinnunen2018tDCF, todisco2019asvspoof} and decrease the gap between the EERs of the evaluation set and development set. 
% , beating the baseline given in \cite{todisco2019asvspoof} using Constant Q cepstral coefficients (CQCC) \cite{todisco2017cqcc} features. 

\begin{table}[t]
\caption{ASV countermeasure-based evaluation}
% \vspace{-0.6cm}
\label{tab: ap-sp}
\vskip 0.5 \baselineskip
\begin{center}
% \begin{small}
% \resizebox{\columnwidth}{!}
\scalebox{0.8}{ \begin{tabular}{l | cc | cc}
\toprule
\multicolumn{1}{c}{} & \multicolumn{2}{c}{Countermeasure EER\%} & \multicolumn{2}{c}{t\_DCF} \\

\hline

\multicolumn{1}{c}{Features} & \multicolumn{1}{c}{DEV} & \multicolumn{1}{c}{EVAL} & \multicolumn{1}{c}{DEV} & \multicolumn{1}{c}{EVAL} \\
\hline
\midrule

\begin{tabular}{@{}c@{}@{}c@{}c@{}} Aperiodic parameters (AP) \\ Spectral envelope (SP) \\MFCC \\ CQCC \\ Spectrogram \\ AP+SP \\AP+SP+MFCC \\ AP+SP+CQCC\\ AP+SP+Spectrogram  \\ \end{tabular} 

& \begin{tabular}{@{}c@{}} 21.19 \\ 10.55\\ 7.14\\ 1.37 \\ 0.48 \\ 9.41 \\ 5.14 \\ 3.85 \\ 0.62   \end{tabular} 

& \begin{tabular}{@{}c@{}} 20.65 \\ 9.31 \\ 11.64\\ 10.89 \\  9.39 \\ 8.91 \\ 8.48 \\ 6.73 \\ 6.67  \end{tabular} 

& \begin{tabular}{@{}c@{}} 0.4374 \\ 0.3520 \\ 0.1942 \\ 0.0407 \\ 0.0132 \\ 0.2872 \\ 0.1560 \\ 0.1293 \\ 0.0201   \end{tabular} 

& \begin{tabular}{@{}c@{}} 0.4445 \\ 0.2453 \\ 0.2663 \\ 0.2746 \\ 0.1954 \\ 0.2462 \\ 0.2169 \\ 0.1777 \\ 0.1604   \end{tabular} \\
% spec: epoch 40 

\bottomrule 
\hline
\end{tabular}
}

\end{center}
\vspace{-0.4cm}
\end{table}

\begin{figure}[t]
\centering
\subfigure{ASV dataset} {\includegraphics[width=\linewidth]{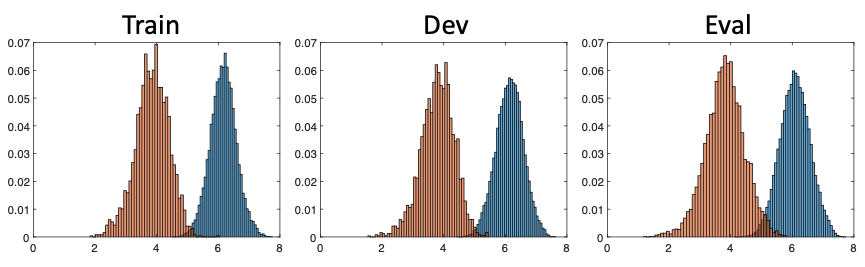} \label{fig:asv}}%
\subfigure{FoR dataset} {\includegraphics[width=\linewidth]{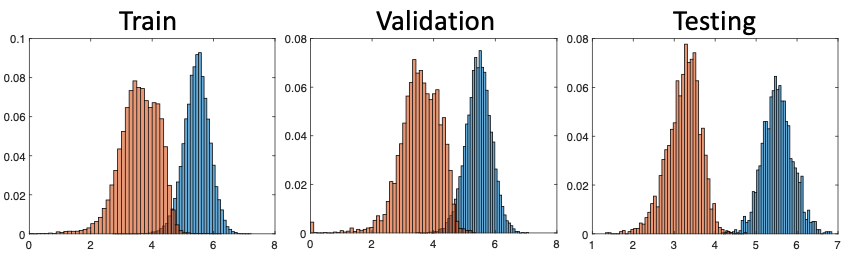} \label{fig:FoR}}
\vspace{-0.5cm}
\caption{Spectral entropy distributions. Blue is for fake speech and orange is for real speech}
\label{fig:spectralE}
\vspace{-0.3cm}
\end{figure}

%\vspace{-0.2cm}

In the case of spectral entropy of F0 sequence, our hypothesis is that the F0 sequence of synthetic speech may lack the characteristic shift and variation of natural speech. We use the Shannon entropy of the power spectral density of the F0 sequence to capture this. The equations for the computation of this spectral density are as equations \eqref{eq:e1}, \eqref{eq:e2} and \eqref{eq:e3}, which first calculate the power spectral density (PSD) of your signal's spectrum $X(w_i)$; then normalize the PSD as probability density function; and finally compute the power spectral entropy. 

\begin{equation}
    P(w_i) = \frac{1}{N}|X(w_i)|^2
    \label{eq:e1}
\end{equation}

\begin{equation}
    P_i = \frac{P(w_i)}{\sum_i P(w_i)}
    \label{eq:e2}
\end{equation}

\begin{equation}
    PSE = -\sum^n p_i\ln{p_i}
    \label{eq:e3}
\end{equation}

The F0 sequence is extracted using WORLD \cite{morise2016world}, and is trimmed to remove the zero values at the beginning and end of the sequence. We plot the spectral entropy distributions for the ASVspoof 2019 logical data's train/dev/eval set and find consistent patterns in them. To evaluate the stability/significance of the patterns, we also compute the distribution from the FoR dataset, as shown in Fig. \ref{fig:spectralE}. Results show that the spectral entropy of F0 sequence is a surprisingly good indicator that captures statistical differences between synthetic speech and natural speech across datasets. 
% The code for the computation of spectral entropy of F0 sequence is released in \cite{code_se}. 

\section{Discussions}
\label{sec:Discussion}

To further understand the anti-spoofing properties of the aperiodic signal and spectral envelope signals, we evaluated their performance with the direct usage in the ASV model. As shown in Table \ref{tab:impersonation-asv-eer} and Table \ref{tab:asv-eer}, AP/SP-based black-boxes and white-boxes show much larger ASV EER$\%$ than STFT/MFCC based manners under most attacks. This is even much more obvious in SP-based boxes. The potential speculation is that both AP and SP are the features corresponding to identity-independent attributes like content-dependent attributes. SP is even mostly disentangled from speaker identities. These results are expected since the AP and SP signals are chosen to capture the nuances differences between natural speech and fake speech, while ASV systems requires features that distinguish the speakers' voice characteristics in a finer level. Still, one interesting phenomena we noticed is that AP/SP features, especially AP, seems to be good supplementary information that gives lower EER$\%$ for attacks that STFT/MFCC are not good at. This is consistent with our hypothesis that they could capture the signature information to distinguish human-generated speech and machine-generated speech.

\section{Conclusions}

% \vspace{-0.2cm}
In this study, we have established that spoofing attacks carried out using deep-fake speech are more likely to be effective than those using other synthetic methods or human impersonation; even the speech is produced by professional impersonators. We have also established that features that capture the fine-level inconsistencies and nuances of the speech production process could consistently exhibit differences between synthetic speech and genuine speech. All of these result in more robust detection of spoofed speech, and result in rendering ASV systems more robust to attacks generated using unseen methods.

\bibliographystyle{IEEEbib}
\bibliography{strings,refs}

\end{document}